# Optical Synoptic Telescopes: New Science Frontiers[*]

J. Anthony Tyson
Physics Dept., University of California, One Shields Ave., Davis, CA USA 95616

## ABSTRACT

Over the past decade, sky surveys such as the Sloan Digital Sky Survey (SDSS) have proven the power of large data sets for answering fundamental astrophysical questions. This observational progress, based on a synergy of advances in telescope construction, detectors, and information technology, has had a dramatic impact on nearly all fields of astronomy, and areas of fundamental physics. The next-generation instruments, and the surveys that will be made with them, will maintain this revolutionary progress. The hardware and computational technical challenges and the exciting science opportunities are attracting scientists and engineers from astronomy, optics, low-light-level detectors, high-energy physics, statistics, and computer science. The history of astronomy has taught us repeatedly that there are surprises whenever we view the sky in a new way. This will be particularly true of discoveries emerging from a new generation of sky surveys. Imaging data from large ground-based active optics telescopes with sufficient étendue can address many scientific missions simultaneously. These new investigations will rely on the statistical precision obtainable with billions of objects. For the first time, the full sky will be surveyed deep and fast, opening a new window on a universe of faint moving and distant exploding objects as well as unraveling the mystery of dark energy.

## 1. INTRODUCTION

Driven by the availability of new instrumentation, there has been an evolution in astronomical science towards comprehensive investigations of new phenomena. Major advances in our understanding of the Universe over the history of astronomy have often arisen from dramatic improvements in our capability to observe the sky to greater depth, in previously unexplored wavebands, with higher precision, or with improved spatial, spectral, or temporal resolution. Substantial progress in the important scientific problems of the next decade (determining the nature of dark energy and dark matter, studying the evolution of galaxies and the structure of our own Milky Way, opening up the time domain to discover faint variable objects, and mapping both the inner and outer Solar System) all require wide-field repeated deep imaging of the sky in optical bands.

## 2. EVOLVING RESEARCH FRONTIER

Large scale sky surveys, such as SDSS, 2MASS, GALEX and many others have proven the power of large data sets for answering fundamental astrophysical questions. This observational progress, based on advances in telescope construction, detectors, and information technology, has had a dramatic impact on nearly all fields of astronomy, and areas of fundamental physics. A new generation of digital sky surveys described below builds on the experience of these surveys and addresses the broad scientific goals of the coming decade.

Astronomy survey science tends to fall into several broad categories: (1) Statistical astronomy, where large datasets of uniformly selected objects are used to determine distributions of various physical or observational characteristics; (2) Searches for rare and unanticipated objects -- every major survey that has broken new ground in sensitivity, sky coverage or wavelength has made important serendipitous discoveries, and surveys should be designed to optimize the chances of finding the unexpected; and (3) Surveys of the sky become a legacy archive for future generations, allowing astronomers interested in a given area of sky to ask what is already known about the objects there, to photometrically or astrometrically calibrate a field, or to select a sample of objects with some specific properties.

---

[*] Plenary talk at SPIE conference on *Ground-based and Airborne Telescopes III*, 28 June, 2010

## 2.1. Surveys and Moore's Law

Survey telescopes have been an engine for discoveries throughout the modern history of astronomy, and have been among the most highly cited and scientifically productive observing facilities in recent years. This observational progress has been based on advances in telescope construction, detectors, and above all, information technology.

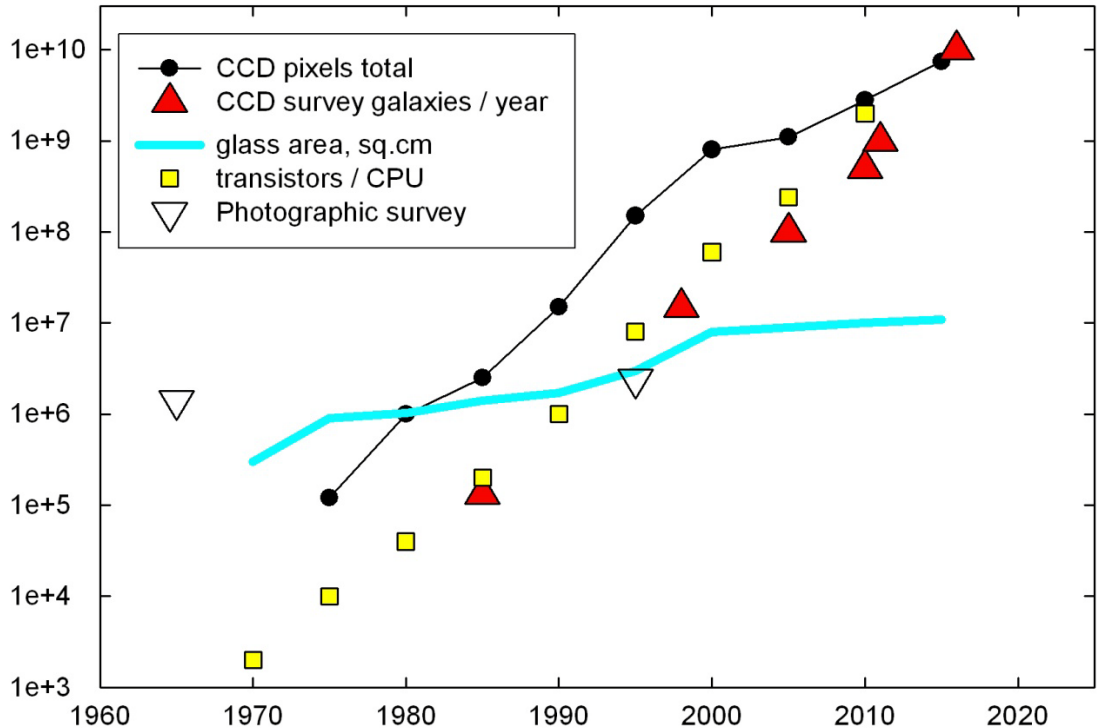

Figure 1 Data trends in optical surveys of the sky. While photographic surveys covered large area, the data were not as usable as digital data and did not go as faint. Information content (in galaxies surveyed per unit time to a given S/N ratio) in CCD digital surveys roughly follows Moore's law. Processing capability has kept up with pixel count. The most recent survey will scan the sky 100 times faster than the 2000 era survey. These next generation wide-fast-deep surveys will open the time window on the universe.

Aided by rapid progress in microelectronics, current sky surveys are again changing the way we view and study the Universe. The next-generation instruments, and the surveys that will be made with them, will maintain this revolutionary progress. Figure 1 charts the trend in optical sky surveys over 50 years. The effect of technology is clear. While the total light collecting area of telescopes has remained comparatively constant, the information content of sky surveys (using the number of galaxies measured per year to a given signal-to-noise ratio as a proxy) has risen exponentially. This is in large part due to high efficiency imaging arrays growing to fill the available focal plane, and to the increase in processing power to handle the data flood. Development of software to analyze these digital data in new ways has resulted in a corresponding increase in science output.

Photographic surveys had the advantage of large focal plane area early on, but have been eclipsed by more sensitive CCD surveys, driven by the exponential rise in pixel count and computer processing capability – both enabled by the microelectronics "Moore's Law". Plotted vs time is the sum of all CCD pixels on the sky, as well as the number of transistors in a typical CPU. Processing capability keeps up with the data rate. Also plotted is one result of CCD surveys – the number of galaxies photometered per unit time – ranging from a survey using a single 160 Kpixel CCD on a 4m telescope to a 3.2 Gpixel camera on an 8.4m telescope. From the start, CCD detectors had more than an order of magnitude higher sensitivity than photographic plates (and far better dynamic range). The population of faint blue galaxies was discovered immediately in the first small CCD survey in the early 1980s. More recently, this high efficiency coupled with many square degree focal plane arrays on large aperture wide-field telescopes means that we can tile the sky quickly with deep exposures -- opening up a new window on the universe.

## 2.2. Astro Sociology

The exponential increase in survey data and the resulting science opportunities has resulted in the development of a new breed of scientist, the "survey astronomer". These include both the people who develop the infrastructure of these surveys and those who analyze these data. The hardware and computational challenges and the exciting science opportunities are attracting scientists from high-energy physics, statistics, and computer science. The way astronomers pursue their science is also evolving. Breakthroughs in observational astronomy in the last fifty years have been driven by two very different types of facilities (often working together):

- *Survey facilities* are usually dedicated telescopes and cameras with a wide field of view, which gather data on large numbers of objects, for use in a wide variety of scientific investigations.

- *Observatories* are designed to allow detailed studies of individual objects or relatively small fields of view in a given waveband. Much of the push towards telescopes of ever larger aperture is motivated by studies of individual objects. The most common are large telescopes, each having a number of interchangeable instruments.

A recent study of the astronomical literature citing telescope facilities found that the SDSS topped the list, ahead of the Hubble Space Telescope (a facility costing 60 times as much).[1] Wide-field and narrow-field facilities alike are driven by new technological developments. The history of astronomy has taught us repeatedly that there are unanticipated surprises whenever we view the sky in a new way. Complete, unbiased surveys are the best technique we have both for discovering new and unexpected phenomena, and for deriving the intrinsic properties of source classes so that their underlying physics can be deduced. As was the case with the Palomar Schmidt and 200-inch, the new digital sky surveys will fuel much of the E-ELT, GMT, and TMT observing.

## 3. NEW DIGITAL SKY SURVEYS

Today there is a new crop of exciting sky survey facilities just starting operation or under construction. The success of the SDSS and other surveys, and the important scientific questions facing us today, have motivated astronomers to plan the next generation of major surveys. In addition to several ground-based projects, there are proposals for complementary space-based surveys at wavelengths unavailable from the ground (WISE is a currently operating example.) In the optical, wide-field cameras are being built for a variety of telescopes. Some of these will share the telescope with other instruments, limiting the time available for surveys. Others are specifically built wide-field telescopes and cameras dedicated to survey work. Some of these surveys will go appreciably deeper than SDSS and with better image quality, and will open the time domain to study the variable universe. Below is a short summary of four new dedicated survey facilities.

### 3.1. SkyMapper

SkyMapper is a 1.35m modified Cassegrain telescope with f/4.8 optics, sited at Siding Spring Observatory near Coonabarabran, NSW Australia.[2] The telescope with an effective 5.2 deg$^2$ field of view 268 megapixel CCD camera will survey the southern sky: about one billion stars and galaxies. Its one-minute exposures will tile each part of the sky 36 times in multiple optical bands, ultimately creating 500 TB of image data. This will be the first comprehensive digital survey of the entire southern sky. A distilled version of the SkyMapper Survey will be made publicly available and will include a set of images of all the stars, galaxies, and nebulae, as well as a database containing the accurate color, position, brightness, variability, and shape of over a billion objects in the southern sky. SkyMapper has an étendue of 5.2 m$^2$deg$^2$ and is now taking commissioning data.

### 3.2. VISTA

VISTA is a 4-m wide field near-IR survey telescope for the southern hemisphere, located at ESO's Cerro Paranal Observatory in Chile.[3] It is equipped with a near infrared camera with 16 hybrid CMOS HgCdTe arrays totaling 67 million pixels of size 0.34 arcsec (0.6 square degree coverage per exposure) and five broad band filters covering 0.85 – 2.3 micron, and a narrow band filter at 1.18 micron. The telescope has an alt-azimuth mount and R-C optics with a fast f/1 primary mirror giving an f/3.25 focus to the instrument at Cassegrain. The site, telescope aperture, wide field, and high quantum efficiency detectors will make VISTA the world's premier ground based near-IR survey instrument. VISTA has an étendue of 6.8 m$^2$deg$^2$ and is now taking science verification survey data.

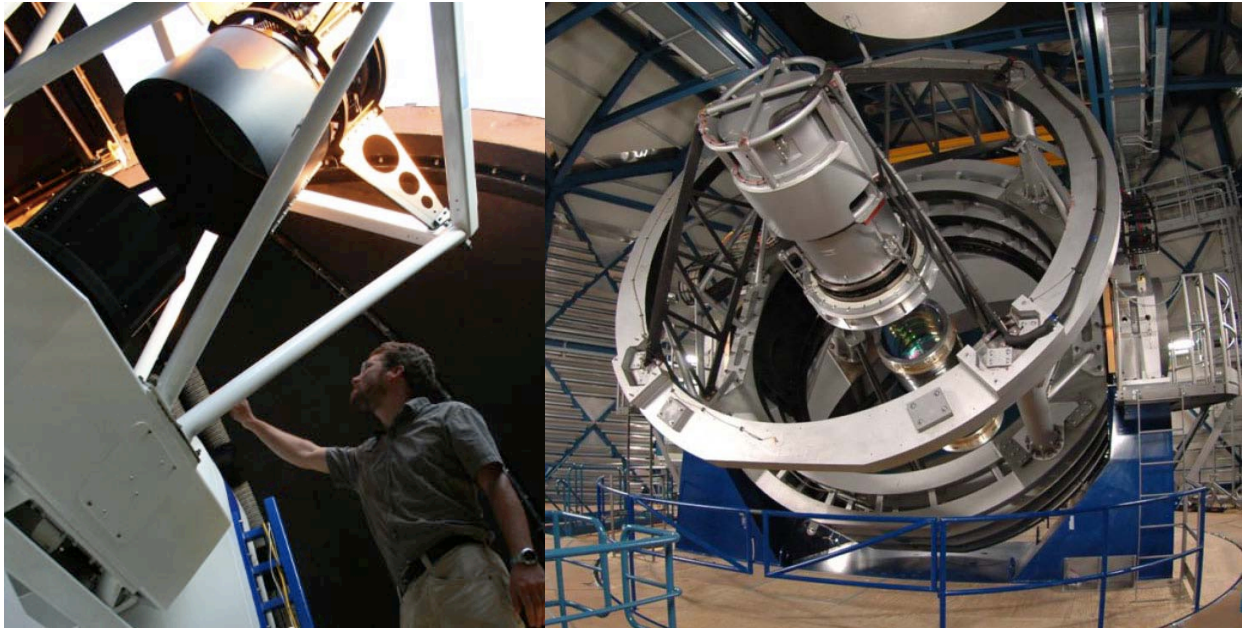

Figure 2. *Left:* SkyMapper 1.4m optical survey telescope with 268 Mpixel CCD camera in Australia. *Right:* VISTA 4m near-IR survey telescope and 67 Mpixel HgCdTe hybrid CMOS imaging array in Chile.

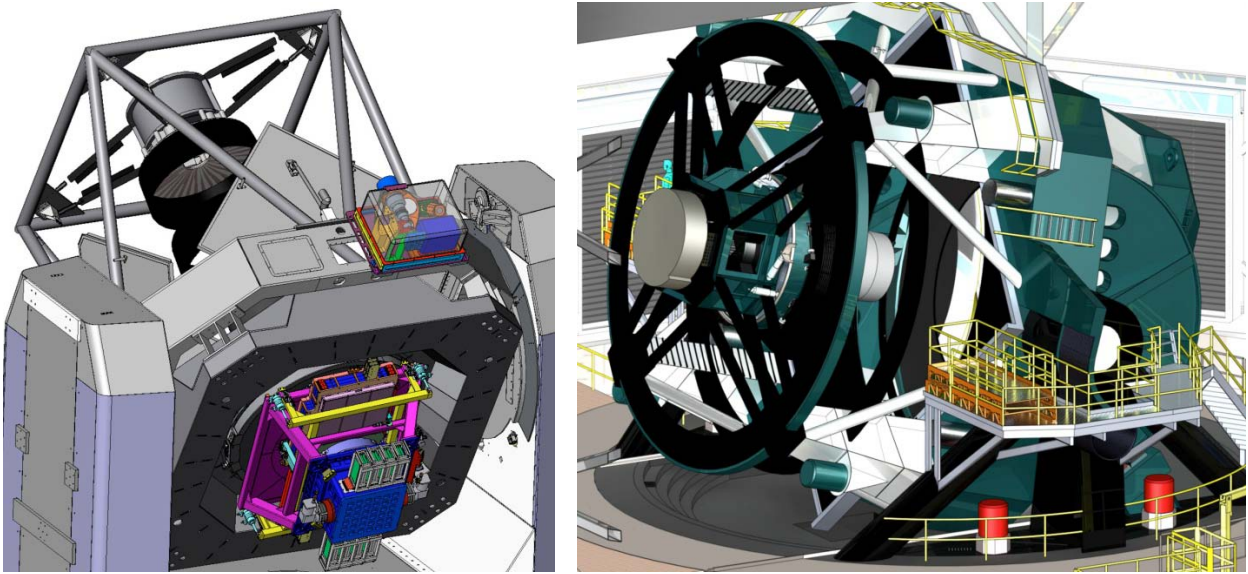

Figure 3. *Left:* Pan-STARRS PS1 1.8m optical survey telescope with 1.4 Gpixel camera in Hawaii. *Right:* LSST 8.4m optical survey telescope with 3.2 Gpixel camera planned for Chile.

### 3.3. Pan-STARRS

The basis of Pan-STARRS is a 1.8m f/4 R-C telescope and 1.4 Gpixel CCD camera in Hawaii.[4] The camera focal plane contains a 64 x 64 array of "orthogonal transfer" CCDs, each containing approximately 600 x 600 pixels, for a total of about 1.4 gigapixels. The planned surveys will use five wide bands spanning 0.4-1.0 micron. PS1 has an étendue of 13 $m^2 deg^2$. The primary science driver is a 30,000 $deg^2$ northern hemisphere survey of near-Earth objects of 1km and even smaller size. However, many other science applications become possible with such a multiband optical survey of the sky. The data will remain proprietary for some time. Ultimately, the plan is to build four such 1.8m facilities. The pilot system PS1 has been built and is undergoing system verification tests and commissioning.

### 3.4. LSST

With an étendue of 319 m$^2$deg$^2$ the 3-mirror Large Synoptic Survey Telescope[5] (LSST) will produce a 6-band (0.3-1.1 micron) wide-field deep astronomical survey of over 30,000 deg$^2$ of the sky using an 8.4m (effective aperture 6.7m) active optics telescope and 3.2-Gpixel camera in Chile. Each patch of sky will be visited 1000 times (2x15 sec exposures each time) in ten years. Twenty trillion photometric and temporal measurements covering 20 billion detected objects will be obtained. The 30 terabytes of pipeline processed data obtained each night will yield a wide-fast survey of the deep optical universe for variability and motion, including asteroids. The deep coverage of ten billion galaxies will provide unique capabilities for cosmology. Astrometry, 6-band photometry, and time domain data on 10 billion stars will enable studies of galactic structure. All data will be open to US and Chile.

### 3.5. Étendue comparison

To survey a statistically significant portion of the universe, a facility must image deeply in a short time and tile a broad swath of sky many times. The rate which a given facility can survey a volume of the universe is a product of its light gathering power (effective aperture in square meters) and the area of the sky imaged in a single exposure (in square degrees). This product is the étendue. Figure 4 compares the étendue of the modern wide-field synoptic digital sky surveys, on a log scale. For good pixel sampling of the PSF, the étendue is also a measure of the data rate. Effective obscured aperture and active imaging field of view are used to calculate the étendue. For an all available sky survey lasting a given number of years, higher étendue translates into more revisits to the same patch of sky and greater depth in the co-added image. Six of the facilities are dedicated to imaging surveys. The remaining share their wide-field cameras with other instruments (estimated fraction). Even so, it can be seen that the Hyper Suprime camera on the 8m Subaru telescope, with its superb image quality, could be quite effective.[6] For studies of galaxies, faint limiting surface brightness is key. This is enabled by high étendue. Image quality is also important. The PSF width of the facilities in Figure 4 spans a range of 5 to 1, and the rate of revisit to a sky patch spans a 1000 to 1 range. Figure 5 compares two images of the same small patch of sky at equal S/N ratio: one with the SDSS and the other with the Subaru telescope to the depth expected in the LSST survey after one year.

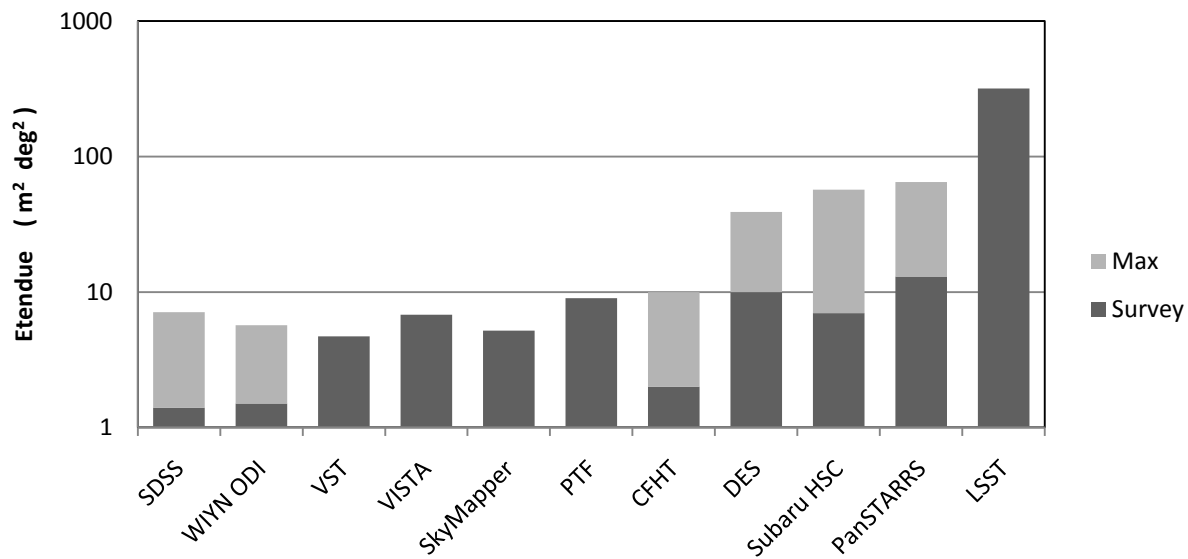

Figure 4. Étendue of current and planned survey telescopes and cameras. Some are dedicated 100% to surveys ("*Survey*"). Others could have higher effective étendue if used 100% in survey mode or if duplicated ("*Max*"). Above an étendue of 200-300 m$^2$deg$^2$ it becomes possible to undertake a single comprehensive multi-band survey of the entire visible sky serving most of the science opportunities, rather than multiple special surveys in series.

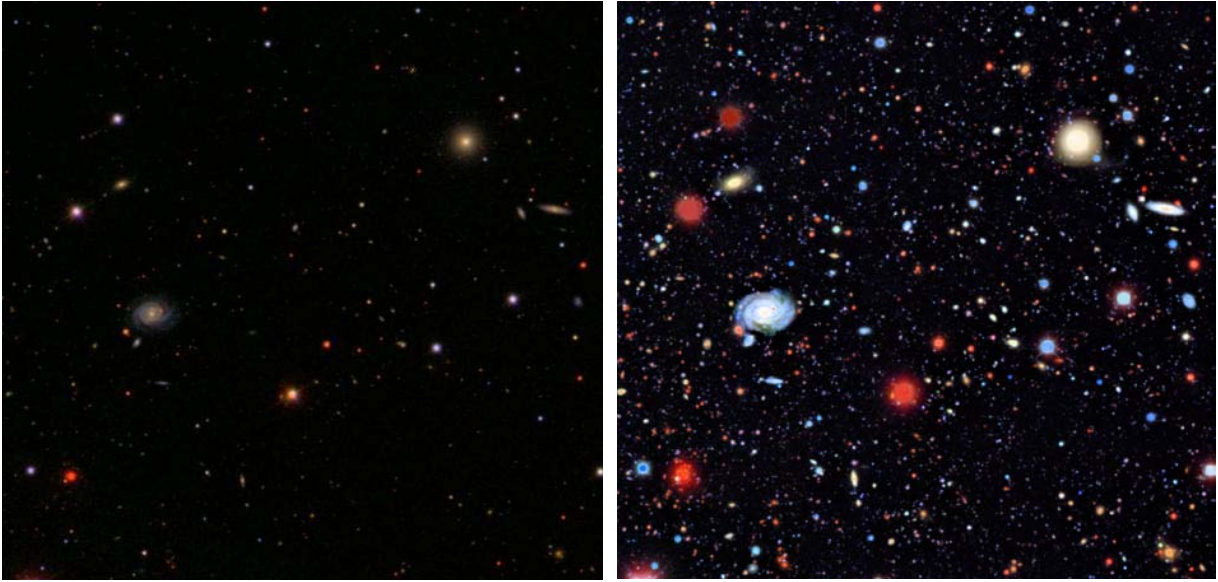

Figure 5. Two views of the sky displayed at the same signal-to-noise. On the *left* is shown a 7.5x7.5 arcminute part of an SDSS field. On the *right*, to LSST one-year depth 2800 galaxies brighter than 25.2 *i* magnitude are seen in this same 0.016 deg$^2$ field [NOAO 4m Deep Lens Survey + Subaru SuprimeCam].

## 4. THE FOURTH DIMENSION

Exploration of the variable optical sky is one of the new observational frontiers in astrophysics. Previous survey catalogs are primarily of the static sky. No optical telescope to date has had the capability to search for transient phenomena at faint levels over enough of the sky to fully characterize phenomena. Variable and transient phenomena have historically led to fundamental insights into subjects ranging from the structure of stars, to the most energetic explosions in the universe, to cosmology. Existing surveys leave large volumes of discovery parameter space (in wavelength, depth, and cadence) unexplored. The exponential increase in survey power shown in Figure 1 translates into an exponential increase in time sampling of the variable universe, adding the time dimension to our view of the cosmos.

Because of the wide coverage and broad time sampling, some of the transient object science will be accomplished largely from the databases resulting from the wide-fast-deep surveys, combined where appropriate with multi-wavelength databases from other facilities, space and ground. For fast or repeating transients, the LSST Deep Drilling sub-survey of ~50 selected ten square degree fields will yield the highest quality data with excellent time sampling. However, much of the transient science enabled by these new surveys will rely on additional observations of selected transient objects based on their classification using the survey photometric data. Some of the additional observations will be in the follow-up mode and some will (optimally) be in a "co-observing" mode where complementary facilities monitor the same sky during operations. Spectroscopic follow-up will be a world effort. A well-designed follow-up strategy must include end-to-end planning and must be in place before first light. In terms of the required photometric and spectroscopic follow-up, generally there are two distinct cases of transients.

*Rare bright transients detected by on their way up:* In this case the transients will be sparse on square degree scales. Efficient follow-up would then focus on one transient at a time. Requirements include multi-band simultaneous photometry and integral field unit (IFU) fiber fed spectroscopy on rapidly deployed telescopes around the world that can continuously follow transients brighter than ~22$^{nd}$ mag. An example is the Las Cumbres Observatory Global Telescope Network of 1m and 2m telescopes and photometric + IFU instruments dedicated to follow-up[7]. It will be important that 1-4m class facilities be capable of following the brief transient to its peak brightness.

*Many faint transients:* Every night LSST is expected to deliver data tens of thousands of astrophysical transients. The majority of these will be moving objects or variable stars. Accurate event classification can be achieved by real-time access to the required context information: multi-color time-resolved photometry and host galaxy information from the survey itself, combined with broad-band spectral properties from external catalogs and alert feeds from other instruments. For photometry, LSST will provide sparsely time-sampled follow-up on hours-days timescales. Because we expect many such transients per field of view, efficient spectroscopic follow-up would best be carried out with multi-slit or multi-IFU systems.

### 4.1. Accurate classification needed

Efficient follow-up will depend on focusing limited resources on the interesting transients. There is always a trade between accuracy and completeness. After the first year of operation LSST will be able to produce enough archival and current transient information to enable accurate event *classification*. Combining the optical transient data with survey or archival data at other wavelengths will be routine through the Virtual Observatory. We expect that the community will make significant progress on classification of transients before LSST operations begin, given lessons learned from Palomar Transient Factory[8] and PS1. Classification accuracy is driven by precision deep photometry and many repeat measurements.

## 5. SCIENCE OPPORTUNITIES AND SURVEY STRATEGY

A wide-fast-deep survey of a large fraction of the sky in multiple optical bands is required in order to explore many of the exciting science opportunities of the next decade. The most important characteristic that determines the speed at which a system can survey the sky to a given depth is its étendue. As shown in Figure 4, the effective étendue of the new generation of sky surveys is rising to more than an order of magnitude larger than that of any existing facility. As was the case with SDSS, we expect the scientific community will produce a rich harvest of discoveries. In coming years this new crop of surveys will make headway in new science directions; we discuss some examples below using LSST science capability studies as an example.

About 90% of the observing time should be devoted to a uniform deep-wide-fast (main) survey mode. The main deep-wide-fast survey mode will observe a 20,000 deg$^2$ region to +15 deg dec about 1000 times (summed over all six bands) with pairs of 15 second exposures during the anticipated 10 years of operations, resulting in a co-added map with 5$\sigma$ point source limiting *r* [650nm] magnitude of 27.7. This map will enable photometric and other measurements of 10 billion stars and a similar number of galaxies. The remaining 10% of the observing time could be allocated to special programs such as a Very Deep + Fast time domain survey (so-called "Deep Drilling" fields). Over 100 science programs are discussed in the LSST Science Book[5]. For transient and variable phenomena LSST will extend time-volume discovery space a thousand times over current surveys. Below we briefly describe three other science themes, with some emphasis on dark energy, in keeping with the theme of these plenary talks.

### 5.1. Inventory of the Solar System

The small bodies of the Solar System offer a unique insight into its early stages. The planned cadence will result in orbital parameters for several million moving objects; these will be dominated by main-belt asteroids, with light curves and colorimetry for a substantial fraction of detected objects. This represents an increase of factors of 10 to 100 over the numbers of objects with documented orbits, colors, and variability information. Ground-based optical surveys are the most efficient tool for comprehensive near-Earth object (NEO) detection, determination of their orbits and subsequent tracking. A survey capable of extending these tasks to Potentially Hazardous Objects (PHAs) with diameters as small as 100 m requires a large telescope, a large field of view and sophisticated data acquisition, processing and dissemination system. A 10 m-class telescope is required to achieve faint detection limits quickly, and together with a large field of view (~10 square degrees), to enable frequent repeated observations of a significant sky fraction -- producing tens of terabytes of imaging data per night. In order to recognize PHAs, determine their orbits and disseminate the results to the interested communities in timely manner, a powerful and fully automated data system is mandatory. A recent NRC study found that a ground-based telescope with an étendue above 300 m$^2$deg$^2$ will be the most cost effective way of achieving the Congressional mandate of detecting and determining the orbits of 90% of potentially hazardous asteroids larger than 140m.[9]

### 5.2. Mapping the Milky Way

The new generation sky surveys are ideally suited to answering two basic questions about the Milky Way Galaxy: What is the structure and accretion history of the Milky Way? What are the fundamental properties of all the stars within 300 pc of the Sun? LSST will enable studies of the distribution of numerous main-sequence stars beyond the presumed edge of the Galaxy's halo, their metallicity distribution throughout most of the halo, and their kinematics beyond the thick disk/halo boundary, and will obtain direct distance measurements below the hydrogen-burning limit for a representative thin-disk sample of stars. For example, LSST will detect of the order $10^{10}$ stars, with sufficient signal-to-noise ratio to enable accurate light curves, geometric parallax and proper motion measurements for about a billion stars. Accurate multi-color photometry can be used for source classification and measurement of detailed stellar properties such as effective temperature and metallicity.

### 5.3. Dark Energy

Dark energy affects two things: distances and the growth of mass structure. We would like to measure the way the expansion of our universe and the mass structure changes with cosmic time. For a universe with a given mass density, the time history of the expansion encodes information on the amount and nature of dark energy. Since the expansion of the universe has been accelerating, the development of mass structures via ordinary gravitational in-fall will be impeded. Measuring how dark matter structures and ratios of distances grow with cosmic time -- via weak gravitational lensing observations -- will provide clues to the nature of dark energy. A key strength of the next generation surveys is the ability to survey huge volumes of the universe. Such a probe will be a natural part of the all-sky imaging survey: billions of distant galaxies will have their shapes and colors measured. Sufficient color data will be obtained for an estimate of the distance to each galaxy by using photometric redshifts derived from the multi-band photometry. Due to its wide coverage of the sky, LSST is uniquely capable of detecting any variation in the dark energy with direction.[10] In turn, this will tell us something about physics at the earliest moments of our universe, setting the course for its future evolution. By combining with dynamical data, we can test whether the "dark energy" is due to a breakdown of General Relativity on large scales.[11]

The LSST will enable scientists to study the dark energy in four different and complementary ways:

1. The telescope will tomographically image dark matter over cosmic time, via a "gravitational mirage." All the galaxies behind a clump of dark matter are deflected to a new place in the sky, causing their images to be distorted. This is effectively 3-D mass tomography of the universe.
2. Galaxies clump in a non-random way, guided by the natural scale that was imprinted in the fireball of the Big Bang. This angular scale [so called "baryonic acoustic oscillations" BAO] will be measured over cosmic time, yielding valuable information on the changing Hubble expansion.
3. The numbers of huge clusters of dark matter are a diagnostic of the underlying cosmology. Charting the numbers of these (via their gravitational mirage: weak lensing, WL) over cosmic time, will place another sensitive constraint on the physics of dark energy.
4. Finally, a million supernovae will be monitored, giving yet another complementary view of the history of the Hubble expansion.

Dark energy affects the cosmic history of the Hubble expansion as well as the cosmic history of mass clustering (which is suppressed at epochs when dark energy dominates). If combined, different types of probes of the expansion history (via distance measures) and dark matter structure history can lead to percent level precision in dark energy parameters. Using the cosmic microwave background as normalization, the combination of these deep probes over wide area will yield the needed precision to distinguish between models of dark energy, with cross checks to control systematic error. Hence, it is desirable for future surveys to provide results of both the distance and growth of dark matter structure, so that different theoretical models can be easily and uniformly confronted with the data. Figure 6 demonstrates for a 20,000 deg$^2$ very deep survey that joint BAO and WL can achieve ~0.5% precision on the distance and ~2% on the structure growth factor over a wide range of redshift. Such measurements can test the consistency of dark energy or modified gravity models.[12] For example, some dynamical dark energy models allow large changes or oscillations in the distance-redshift relation over this range of redshifts – these models can then be tested.

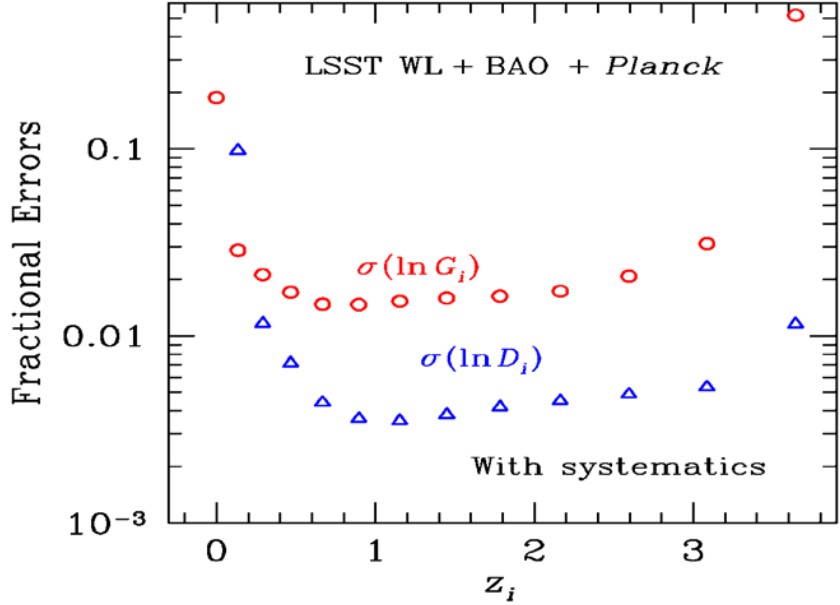

Figure 6. Joint baryon acoustic oscillation and weak gravitational lensing constraints on the comoving distance (triangles) and dark matter structure growth factor (circles) from a 20,000 deg$^2$ deep survey in six optical bands. These data will enable percent level tests of theories of dark energy, whether modified gravity or stress-energy.

## 6. FROM SCIENCE DRIVERS TO SYSTEM SPECIFICATIONS

Given the science opportunities listed in the previous section, we now review how they are translated into survey operations requirements and constraints on the main system design parameters: the aperture size, the survey lifetime, and the optimal exposure time. We use LSST science-instrument flow-down as an example. The field-of-view area is set to the practical limit possible with modern optical designs, 10 deg$^2$, determined by the requirement that the delivered image quality be dominated by atmospheric seeing at the chosen site. A larger field-of-view would lead to unacceptable deterioration of the image quality. This leaves the primary mirror diameter and survey lifetime as free parameters. Our adopted survey lifetime is ten years. Shorter than this would imply an excessively large and expensive mirror (15 meters for a three-year survey and 12 meters for a five-year survey), while a much smaller telescope would require much more time to complete the survey with the associated increase in operations cost and evolution of the science goals. The primary mirror size is a function of the required survey depth and the desired sky coverage. Roughly speaking, the anticipated science outcome scales with the number of detected sources. For practically all astronomical source populations, in order to maximize the number of detected sources, it is more advantageous to maximize first the area and then the detection depth.

- *The single visit depth* should reach $r$ [650 nm] = 24.7 magnitude (5$\sigma$ point source). This limit is driven by need to image faint moving, variable, or transient sources, and by proper motion and trigonometric parallax measurements for stars. Indirectly, it is also driven by the requirements on the co-added survey depth and the minimum number of exposures required by weak lensing science to average over systematics in the point-spread function.

- *Image quality* should maintain the limit set by the atmosphere (for free-air PSF of 0.4-0.6 arcsec FWHM at 600-800nm) and not be degraded appreciably by the hardware. This requirement comes from weak lensing, survey depth for point sources, and image differencing for detection of moving and transient sources. The system contribution to the PSF must be under 0.3 arcsec FWHM.

- *Photometric repeatability* should achieve half percent precision at the bright end, with photometric zeropoint stability across the sky of one percent and band-to-band calibration errors not larger than half percent. These requirements are driven by the need for photometric redshift accuracy, the separation of stellar populations, detection of low-amplitude variable objects (such as eclipsing planetary systems), and the search for systematic effects in supernova light curves.

- *Astrometric precision* should maintain the limit set by the atmosphere of about 10 milliarcseconds (mas) rms per coordinate per visit at the bright end on scales below 20 arcmin. This precision is driven by the goal of proper motion uncertainty of 0.2 mas/yr and parallax uncertainty of 1.0 mas over the course of a 10-year survey.

- *The single visit total exposure time* (summed over two exposures in a visit) should be about 30 sec to prevent trailing of fast moving objects and to aid control of various systematic effects induced by the atmosphere. It should be longer than ~20 seconds to avoid significant efficiency losses due to finite readout, slew time, and CCD read noise.

- *The filter complement* should include six filters in the wavelength range limited by atmospheric absorption and silicon detection efficiency (320--1080 nm), with roughly rectangular passbands and no large gaps in the coverage, in order to enable robust and accurate photometric redshifts and stellar typing. A 400nm band is extremely important for separating low-redshift quasars from hot stars and for stellar studies. A filter with an effective wavelength of 1 micron will enable photometric redshifts, studies of sub-stellar objects, high-redshift quasars, and regions of the Galaxy that are obscured by interstellar dust. All six bands are required in order to meet photometric redshift science requirements.

- *The revisit time distribution* should enable determination of orbits of solar system objects and sample supernova light curves every few days, while accommodating constraints set by proper motion and trigonometric parallax measurements.

- *The total number of 2-exposure visits* of any given area of sky, when accounting for all filters, should be about 1,000, as mandated by weak lensing science, the asteroid survey, and proper motion and trigonometric parallax measurements. Studies of variable and transient sources of all sorts also require a large number of visits.

- *The coadded survey depth* should reach $r \sim 27.7$ magnitude ($5\sigma$, point source), with sufficient signal-to-noise ratio in other bands to address both extragalactic and Galactic science drivers. $S/N > 10$ for galaxies brighter than 26 $r$ mag is required for photometric redshifts, so flexibility of this limit is only ~0.3 mag.

- *The distribution of visits on the sky* should extend over at least 20,000 deg$^2$ to obtain the required number of galaxies for cosmology studies, to study the distribution of galaxies on the largest scales and to probe the structure of the Milky Way and the solar system, with attention paid to include special regions such as the ecliptic, the Galactic plane, and the Large and Small Magellanic Clouds.

- *Data processing, data products and data access* should enable efficient science analysis without a significant impact on the final uncertainties. To enable a fast and efficient response to transient sources, the processing latency for objects that change should be less than a minute after the close of the shutter, with a robust and accurate preliminary classification of reported transients.

A survey facility with étendue $> 300$ m$^2$deg$^2$ will meet the requirements for these plus a very broad range of other scientific programs. Such a system can adopt a highly efficient survey strategy where *a single dataset serves most science programs* (instead of science-specific surveys executed in series). The science requirements on co-added depth and the survey sky coverage drive the aperture, the field-of-view size, and the survey length. With the sky coverage and the field of view size set to their practical limits, the co-added depth requirements yield an effective aperture of at least 6.5m for a 10 year survey (the LSST effective aperture is 6.7 m). This depth requirement is based on detailed simulations using the LSST Operations Simulator. With the aperture determined by the final co-added depth requirements, the system deployment parameters, such as exposure time, are driven by numerous additional constraints derived mostly from time sampling requirements (see Figure 7). One of the most important lessons of the previous generation of surveys is that the interesting science questions at the end of the survey can be different than they were when the surveys were being planned. These surveys succeeded in this evolving terrain by being very general tools that could be applied to a number of very fundamental measurements.

Likewise, the public availability and accessibility of the data enabled the broader astronomical community to generate more science than the survey collaborations could alone. An open-source open-data policy benefits all and ensures innovation in applications software. Therefore a survey should plan to make its data and software pipelines public and properly documented in a form that allows the full scientific community to use them.[13]

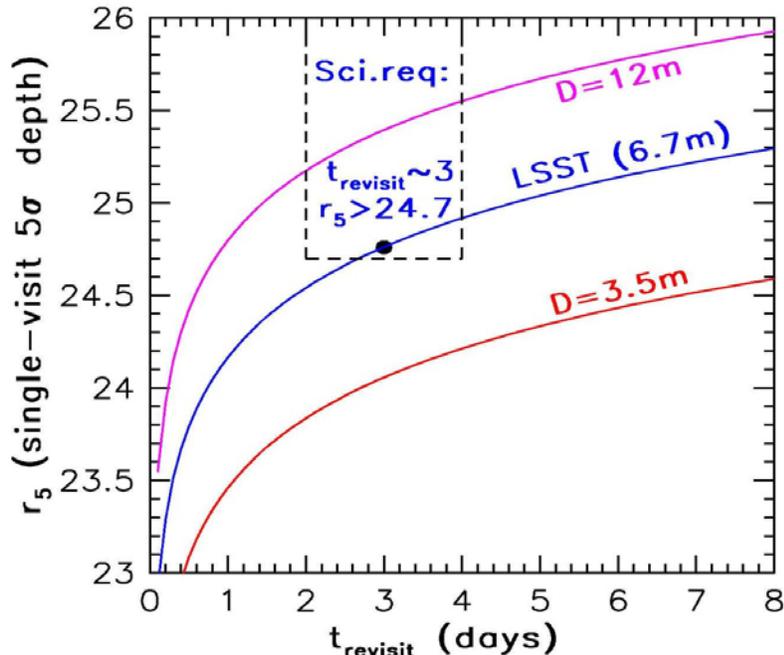

Figure 7. For a 10 deg$^2$ field, the science requirements lead to an étendue requirement for a single 6-band comprehensive survey. Shown is the single-visit depth at 650nm in magnitudes vs revisit time for a 10 year survey covering a hemisphere. Three effective aperture curves are plotted at étendue 100, 320, and 1100 m$^2$deg$^2$. Multiple constraints result in a fairly narrow exposure time range per visit for a wide-deep-fast survey, and a minimum étendue.

This requires sophisticated databases and extensive documentation, which must be budgeted when the survey is first designed. Data quality is paramount. A lesson that SDSS and other surveys learned, is that it is cheaper to do things right the first time. Survey requirements are defined by first deciding on the core science goals, then designing a telescope, instrument and survey strategy that will meet these goals. Then ask what the data quality that these, together with the laws of nature allow, and design rigorous quality validation tools. This will yield a much more uniform dataset and enable science well beyond that anticipated at the time the survey was designed.

## 7. PETASCALE DATA-TO-KNOWLEDGE CHALLENGE

Petabyte databases are rare today, but will soon be commonplace. Datasets are growing exponentially, a result of the exponential growth of the underlying technologies which allow us to gather the data. Naively, one might think that these datasets could be processed and analyzed in constant time, given the same exponential growth of computing power. However, this is true only for linear algorithms, and only to the extent that all processing and analysis can be automated. Moreover, most databases are collected for a single purpose. Until now analysis of these databases has focused on discovery of predetermined features. This is true over a broad spectrum of applications, from high energy physics, to astronomy, to geoscience. Optimal filters are built to detect given events or patterns. Most often the database itself is pre-filtered and tuned to a well defined application or result, and automated search algorithms are designed which detect and produce statistics on these pre-determined features. However, paradigm-shifting discoveries in science often are the result of a different process: upon close examination of the data a scientist finds a totally unexpected event or correlation.

Large synoptic imaging projects are a perfect example. A time sequence of images is searched for given features. The unexpected goes undetected. One solution to this failure to detect the unexpected rare event is to build visualization tools which keep the human partially in the loop. This still misses most rare events in huge databases, or the unexpected statistical correlation. In the near future gigapixel cameras will flood us with data which must be processed and analyzed immediately. Automated software pipelines will relentlessly search for changes. Most any kind of change at bright levels will be detected. The problem occurs at the faint end where most of the volume is.

There, we will have matched filters to detect *known* classes of objects (variable stars, supernovae, moving objects). But we cannot build a matched filter for something that is beyond our current imagination. The data space is position, time, intensity, colors, and motion. One challenge is to efficiently find rare events in large databases and gauge their significance. Another challenge is finding unexpected statistical correlations. There will be a need for developing statistically efficient estimation schemes and methods for assessing estimation error. Combining statistical efficiency with computational efficiency will be a constant challenge, since the more statistically accurate estimation methods will often be the most computationally intensive. Such automated data search algorithms will serve multiple purposes, from revealing systematic errors in the data for known features to discovery of unexpected correlations in nature. This approach is pointless unless we couple with advanced visualization. By combining automated search algorithms with multi-dimensional visualization we can speed the process of discovery by harnessing simultaneously the best capabilities of the machine and mind.

It will be particularly effective to have data analysis innovations in place when the next generation surveys are on-line. Achieving this requires parallel efforts on optics, electronics, and software. Visualization techniques will aid the development of these automated algorithms. The collaboration will include astronomers and physicists involved in current and upcoming ultra-large surveys, experts on statistics and algorithm development, computer science, data mining, and visualization.

This work was supported by DOE grant DE-FG02-07ER41505, and a TABASGO Foundation grant.